\documentstyle[twoside,epsf]{article}

\catcode`\@=11
\long\def\@makefntext#1{
\protect\noindent \hbox to 3.2pt {\hskip-.9pt
$^{{\eightrm\@thefnmark}}$\hfil}#1\hfill}       

\def\@makefnmark{\hbox to 0pt{$^{\@thefnmark}$\hss}}    

\def\ps@myheadings{\let\@mkboth\@gobbletwo
\def\@oddhead{\hbox{}
\rightmark\hfil\eightrm\thepage}
\def\@oddfoot{}\def\@evenhead{\eightrm\thepage\hfil
\leftmark\hbox{}}\def\@evenfoot{}
\def\sectionmark##1{}\def\subsectionmark##1{}}

\oddsidemargin=\evensidemargin
\newcounter{sectionc}\newcounter{subsectionc}\newcounter{subsubsectionc}
\renewcommand{\section}[1] {\vspace{12pt}\addtocounter{sectionc}{1}
\setcounter{subsectionc}{0}\setcounter{subsubsectionc}{0}\noindent
    {\tenbf\thesectionc. #1}\par\vspace{5pt}}
\renewcommand{\subsection}[1] {\vspace{12pt}\addtocounter{subsectionc}{1}
\setcounter{subsubsectionc}{0}\noindent
{\bf\thesectionc.\thesubsectionc. {\kern1pt \bfit #1}}\par\vspace{5pt}}
\renewcommand{\subsubsection}[1] {\vspace{12pt}\addtocounter{subsubsectionc}{1}
    \noindent{\tenrm\thesectionc.\thesubsectionc.\thesubsubsectionc.
    {\kern1pt \tenit #1}}\par\vspace{5pt}}
\newcommand{\nonumsection}[1] {\vspace{12pt}\noindent{\tenbf #1}
    \par\vspace{5pt}}

\newcounter{appendixc}
\newcounter{subappendixc}[appendixc]
\newcounter{subsubappendixc}[subappendixc]
\renewcommand{\thesubappendixc}{\Alph{appendixc}.\arabic{subappendixc}}
\renewcommand{\thesubsubappendixc}
    {\Alph{appendixc}.\arabic{subappendixc}.\arabic{subsubappendixc}}

\renewcommand{\appendix}[1] {\vspace{12pt}
        \refstepcounter{appendixc}
        \setcounter{figure}{0}
        \setcounter{table}{0}
        \setcounter{lemma}{0}
        \setcounter{theorem}{0}
        \setcounter{corollary}{0}
        \setcounter{definition}{0}
        \setcounter{equation}{0}
        \renewcommand{\thefigure}{\Alph{appendixc}.\arabic{figure}}
        \renewcommand{\thetable}{\Alph{appendixc}.\arabic{table}}
        \renewcommand{\theappendixc}{\Alph{appendixc}}
        \renewcommand{\thelemma}{\Alph{appendixc}.\arabic{lemma}}
        \renewcommand{\thetheorem}{\Alph{appendixc}.\arabic{theorem}}
        \renewcommand{\thedefinition}{\Alph{appendixc}.\arabic{definition}}
        \renewcommand{\thecorollary}{\Alph{appendixc}.\arabic{corollary}}
        \renewcommand{\theequation}{\Alph{appendixc}.\arabic{equation}}
        \noindent{\tenbf Appendix \theappendixc #1}\par\vspace{5pt}}
\newcommand{\subappendix}[1] {\vspace{12pt}
        \refstepcounter{subappendixc}
        \noindent{\bf Appendix \thesubappendixc. {\kern1pt \bfit #1}}
    \par\vspace{5pt}}
\newcommand{\subsubappendix}[1] {\vspace{12pt}
        \refstepcounter{subsubappendixc}
        \noindent{\rm Appendix \thesubsubappendixc. {\kern1pt \tenit #1}}
    \par\vspace{5pt}}

\newcommand{\textlineskip}{\baselineskip=13pt}
\newcommand{\smalllineskip}{\baselineskip=10pt}

\newcommand{\copyrightheading}[1]
    {\vspace*{-2.5cm}\smalllineskip{\flushleft
    {\footnotesize Quantum Information and Computation, Vol.~10, No.~3\&4 (2010) 0222--0232 #1}\\
    {\footnotesize \copyright\kern2pt Rinton Press}\\
     }}

\def\abstracts#1#2#3{{
    \centering{\begin{minipage}{4.5in}\footnotesize\baselineskip=10pt
    \parindent=0pt #1\par
    \parindent=15pt #2\par
    \parindent=15pt #3
    \end{minipage}}\par}}

\def\keywords#1{{
    \centering{\begin{minipage}{4.5in}\footnotesize\baselineskip=10pt
    {\footnotesize\it Keywords}\/: #1
     \end{minipage}}\par}}

\renewenvironment{thebibliography}[1]
        {\frenchspacing
     \ninerm\baselineskip=11pt
         \begin{list}{\arabic{enumi}.}
        {\usecounter{enumi}\setlength{\parsep}{0pt}
     \setlength{\leftmargin 12.7pt}{\rightmargin 0pt}
         \setlength{\itemsep}{0pt} \settowidth
    {\labelwidth}{#1.}\sloppy}}{\end{list}}

\newcounter{itemlistc}
\newcounter{romanlistc}
\newcounter{alphlistc}
\newcounter{arabiclistc}

\newcommand{\fcaption}[1]{
        \refstepcounter{figure}
        \setbox\@tempboxa = \hbox{\footnotesize Fig.~\thefigure. #1}
        \ifdim \wd\@tempboxa > 5in
           {\begin{center}
        \parbox{5in}{\footnotesize\smalllineskip Fig.~\thefigure. #1}
            \end{center}}
        \else
             {\begin{center}
             {\footnotesize Fig.~\thefigure. #1}
              \end{center}}
        \fi}

\newcommand{\tcaption}[1]{
        \refstepcounter{table}
        \setbox\@tempboxa = \hbox{\footnotesize Table~\thetable. #1}
        \ifdim \wd\@tempboxa > 5in
           {\begin{center}
        \parbox{5in}{\footnotesize\smalllineskip Table~\thetable. #1}
            \end{center}}
        \else
             {\begin{center}
             {\footnotesize Table~\thetable. #1}
              \end{center}}
        \fi}

\def\pmb#1{\setbox0=\hbox{#1}
    \kern-.025em\copy0\kern-\wd0
    \kern.05em\copy0\kern-\wd0
    \kern-.025em\raise.0433em\box0}

\def\fnt#1#2{\footnotetext{\kern-.3em
    {$^{\mbox{\scriptsize #1}}$}{#2}}}

\def\runninghead#1#2{\pagestyle{myheadings}
\markboth{{\protect\footnotesize\it{\quad #1}}\hfill}
{\hfill{\protect\footnotesize\it{#2\quad}}}}
\font\tenrm=cmr10
\font\tenit=cmti10
\font\tenbf=cmbx10
\font\bfit=cmbxti10 at 10pt
\font\ninerm=cmr9

\font\eightrm=cmr8

\def\FigName{figure}%
\newbox\captionbox
\long\def\@makecaption#1#2{%
  \ifx\FigName\@captype
    \vskip\abovecaptionskip
    \setbox\tempbox\hbox{{\figurecaptionfont #1\hskip1em #2}}
    \ifdim\wd\tempbox< 28pc
    \centerline{\box\tempbox}
    \else
    {\figurecaptionfont #1\hskip1em #2\par}
\fi\else
    \setbox\tempbox\hbox{{\tablecaptionfont #1\hskip1em #2}}
    \ifdim\wd\tempbox< 28pc
    \centerline{\box\tempbox}
    \else
    {\tablecaptionfont #1\hskip1em #2\par}%
    \fi
 \vskip\belowcaptionskip
 \fi}
\InputIfFileExists{psfig.sty}
{\typeout{^^Jpsfig.sty inputed...ok}}{\typeout{^^JWarning: psfig.sty could be be found.^^J}}
\InputIfFileExists{epsfsafe.tex}
{\typeout{^^Jepsfsafe.tex inputed...ok}}
            {\typeout{^^JWarning: epsfsafe.tex could not be found.^^J}}
\InputIfFileExists{epsfig.sty}
{\typeout{^^Jepsfig.sty inputed...ok}}{\typeout{^^JWarning: epsfig.sty could not be found.^^J}}
\InputIfFileExists{epsf.sty}
{\typeout{^^Jepsf.sty inputed...ok}}{\typeout{^^JWarning: epsf.sty could not be found.^^J}}%
\def\fps@figure{tbp}
\def\ftype@figure{1}
\def\ext@figure{lof}
\def\fnum@figure{Fig.\ \thefigure}
\def\qed{\hbox{${\vcenter{\vbox{              
   \hrule height 0.4pt\hbox{\vrule width 0.4pt height 6pt
   \kern5pt\vrule width 0.4pt}\hrule height 0.4pt}}}$}}

\begin{document}
\setlength{\textheight}{8.0truein}    

\runninghead{
 Computable constraints on entanglement-sharing of multipartite quantum states}
            {Y.-C. Ou and M.S. Byrd}

\normalsize\textlineskip
\thispagestyle{empty}

\centerline{\bf
COMPUTABLE CONSTRAINTS ON
ENTANGLEMENT-SHARING} \vspace*{0.035truein} \centerline{\bf  OF MULTIPARTITE QUANTUM STATES }
\vspace*{0.37truein} \centerline{\footnotesize
YONG-CHENG OU} \vspace*{0.015truein} \centerline{\footnotesize\it
Physics Department, Southern Illinois University} \baselineskip=10pt
\centerline{\footnotesize\it Carbondale, Illinois 62901-4401, USA}
\vspace*{10pt} \centerline{\footnotesize MARK S. BYRD}
\vspace*{0.015truein} \centerline{\footnotesize\it Physics
Department and Computer Science Department, Southern Illinois
University} \baselineskip=10pt \centerline{\footnotesize\it
Carbondale, Illinois 62901-4401, USA} \vspace*{0.225truein}

\vspace*{0.21truein}
\abstracts{
Negativity is regarded as an important measure of entanglement in
quantum information theory. In contrast to other measures of
entanglement, it is easily computable for bipartite states in
arbitrary dimensions. In this paper, based on the negativity and
realignment, we provide a set of entanglement-sharing constraints
for multipartite states, where the entanglement is not necessarily
limited to bipartite and pure states, thus aiding in the
quantification of constraints for entanglement-sharing. These may
find applications in studying many-body systems.}{}{}

\vspace*{10pt} \keywords{Negativity, Realignment, Monogamy,
Entanglement} \vspace*{3pt} 

\vspace*{1pt}\textlineskip  
\section{Introduction}           
\vspace*{-0.5pt}
\noindent
Entanglement is an important resource for quantum information
processing (QIP) and is generally believed to be a key resource in
quantum algorithms.  Although pure entangled states are highly
desirable in QIP \cite{Ekert:91,Bennett:93}, the available states
are most often mixed due to noises of various types in real
experiments. Consequently, in recent years a great deal of effort
has been made to develop methods for  detecting, quantifying, and
characterizing entanglement of bipartite and multipartite quantum
states \cite{Horodeckis:08,Guhne/Toth:08}.

Detection and characterization of entanglement are most often based
on the concept of separability proposed by Werner \cite{Werner:89}.
However, while some lower-dimensional bipartite states have been
quantified thoroughly, for higher dimensions the Peres-Horodecki
criterion \cite{Peres,Horodeckis} detects entanglement of many
states, but not all. Quantifying entanglement has been even more
difficult. For example, one of the most prevalent measures of
entanglement, the entanglement of formation (EOF), has an analytical
formula only for two-qubit states \cite{Wootters:98}. Finding the
EOF for other systems of states is a challenging and open problem.
More recently new methods for describing entangled states have been
developed such as the realignment criterion \cite{Rudolph:02,
Chen/Wu}, entanglement witnesses \cite{Toth/Guhne:05}, covariance
matrices approach \cite{Guhne/etal:07}, and improved realignment
criterion \cite{Zhang/etal:06}, among others.  From these, several
useful lower bounds for the concurrence
 have been derived
\cite{Zhang,Mintert/etal:04,Chen/etal:05,Breuer:06,deVicente:07,Li:07,Gao/etal:06,Fei/Li:06,Zhang/etal:06,Song/etal:07}.
Some measures of entanglement, though being unable to detect bound
entanglement, are relatively easy to calculate.

Furthermore, the quantification of entanglement of multipartite
states is even more difficult although the important monogamy nature
of entanglement was discovered, which states that the entanglement
between the particle $A$ and $B$ constrains that between the
particle $A$ and $C$. The monogamy inequalities developed so far,
such as those derived by Coffman, Kundu, and Wootters
\cite{Coffman/Kundu/Wootters}, hold only for qubit systems
\cite{Ou:07}.  Also, not all measures of entanglement satisfy this
relation. It turns out that there exists other versions of monogamy
inequalities based on squashed entanglement
\cite{Christandl/Winter:04}, distillable entanglement
\cite{Koashi/Winter:04}, etc. Since most measures of entanglement
are difficult to calculate, analytical bounds are quite desirable.

Recently, we found an analytical lower bound of
concurrence~\cite{Ou:08}, which satisfies a monogamy inequality.  We
also found the monogamy of negativity \cite{Ou/Fan:07}.
Interestingly, for Gaussian states, the monogamy of negativity also
holds \cite{Hiroshima/etal:07} and there exists an even stronger
bound \cite{Adesso/Illiuminati:08}. There was an attempt to use
convex-roof extended negativity \cite{sanders} to define a monogamy
entanglement for three-qubit states, but it is unclear whether such
a property is true for pure tripartite states with arbitrary
dimensions. As for the monogamy of negativity itself
\cite{Ou/Fan:07}, it is not clear either.

In this paper, we will establish a set of much more general
constraints on entanglement-sharing in multipartite states, which
are computable. These complement the constraints given in
Ref.~\cite{Ou/Fan:07}.  Specifically, Section 2 introduces the basic
concepts of monogamy of concurrence, negativity, and realignment.
Section 3 proves the more general monogamy of negativity and
realignment, and computable inequalities for any three-qubit states.
Section 4 provides applications, methods for testing these
inequalities through measurements on sets of qubits, and Section 5
concludes.

\section{Monogamy of entanglement}

\noindent For a mixed state $\rho_{AB}$, one well-known measure of
entanglement is the EOF \cite{Wootters:98} defined by
$E(\rho_{AB})=\min_{\{p_i, |\psi_i\rangle\}}\sum_i
p_iE(|\psi_i\rangle)$ for all possible ensemble realizations
$\rho_{AB}=\sum_ip_i|\psi_i\rangle\langle\psi_i|$. Here $p_i\geq 0$,
$\sum_i p_i=1$, and $E(|\psi_{AB}\rangle)=-\texttt{Tr}\rho_A\log_2
\rho_A$ with $\rho_A=\texttt{Tr}_B
|\psi_{AB}\rangle\langle\psi_{AB}|$.  For two-qubit states, the EOF
can be expressed as a function of concurrence defined by
\cite{Wootters:98} \noindent
\begin{equation}\label{eq:1}
{\mathcal {C}}(\rho)  \equiv \max
\{0,\sqrt{\lambda_{1}}-\sqrt{\lambda_2}-\sqrt{\lambda_3}-\sqrt{\lambda_4}\}
,
\end{equation}
where $\lambda_1,...,\lambda_4$ are the singular values of the
matrix
$\rho(\sigma_y\otimes\sigma_y)\rho^*(\sigma_y\otimes\sigma_y)$ in
nonincreasing order, the notation of $*$ stands for complex
conjugation in a particular basis, and $\sigma_y$ is the standard
Pauli spin matrix. For a pure three-qubit state $\rho_{ABC}$, there
exists an inequality in terms of concurrence which provides a
quantification of three-qubit entanglement
\cite{Coffman/Kundu/Wootters}
\begin{equation}
{{{{\mathcal{C}}}}}_{AB}^{2}+{{{{\mathcal{C}}}}}_{AC}^{2}\leq
{{{{\mathcal{C}}}}}_{A:BC}^{2}, \label{a}
\end{equation}%
where ${{{\mathcal{C}}}}_{AB}$ and ${{{\mathcal{C}}}}_{AC}$ are the
concurrences of the mixed states
$\rho_{AB}=\mathrm{Tr}_{C}(|\phi\rangle_{ABC}\langle\phi|)$ and
$\rho_{AC}=\mathrm{Tr}_{B}(|\phi\rangle_{ABC}%
\langle\phi|)$, respectively, and
${{{\mathcal{C}}}}_{A:BC}=2\sqrt{\det\rho_A}$ with
$\rho_{A}=\mathrm{Tr}_{BC}(|\phi\rangle_{ABC}\langle\phi|)$.  We
refer to this as the CKW inequality \cite{Coffman/Kundu/Wootters}.
Based on this inequality in Eq.~(\ref{a}) the three-tangle is
defined as
\begin{equation} \label{22}
\tau_{ABC}={{{{\mathcal{C}}}}}_{A:BC}^{2}-{{{{\mathcal{C}}}}}_{AB}^{2}-{{{{\mathcal{C}}}}}_{AC}^{2},
\end{equation}
which characterizes three-way entanglement of the state and is an
entanglement monotone \cite{Dur/etal:00}. For example, quantified by
the three-tangle, the state $|GHZ\rangle
=\frac{1}{\sqrt{2}}(|000\rangle+|111\rangle )$ has only three-way
entanglement, while the state $|W\rangle
=\frac{1}{\sqrt{3}}(|100\rangle +|010\rangle +|001\rangle )$ has
only two-way entanglement. For general three-qubit mixed states
$\rho_{ABC}$, the three-tangle is defined as
\begin{equation}
\label{ppp} \tau_{ABC} =
\min\left[{{{{\mathcal{C}}}}}_{A:BC}^{2}\right]
                 -{{{{\mathcal{C}}}}}_{AB}^{2}-{{{{\mathcal{C}}}}}_{AC}^{2},
\end{equation}
where
\begin{equation}\label{eq:minconc}
\min\left[{{{{\mathcal{C}}}}}_{A:BC}^{2}\right] =
         {\min}_{\{p_i,{|\psi_i}\rangle\}}\left[\sum_i p_i
         {{{{\mathcal{C}}}}}_{A:BC}^{2}({|\psi_i\rangle})\right]
\end{equation}
denotes the minimum over all pure-state decompositions such that
$\rho_{A:BC} = \sum p_i  {|\psi_i\rangle}{\langle\psi_i|}$.

Two other important measures of entanglement are the negativity and
the realignment. The negativity is based on trace norm of the
partial transpose $\rho^{T_A}$ of a bipartite state $\rho_{AB}$, and
in fact it is a quantitative version of the positive partial
transposition (PPT) criterion for separability
\cite{Peres,Horodeckis}. For a bipartite state $\rho$ the partial
transpose with respect to subsystem $A$ is
\begin{equation}
(\rho ^{T_{A}})_{ij,kl}=\rho _{kj,il},
\end{equation}
and the negativity is defined by
\begin{equation}
{{{{\mathcal{N}}}}}=\frac{{{\Vert} \rho ^{T_{A}}\Vert -1}} {2},
\end{equation}
where the trace norm $\Vert R\Vert $ is given by $\Vert R\Vert
=\mathtt{Tr}\sqrt{ \emph{{R}} ^{\dagger}\emph{R}}$.  The negativity
can also be seen to be equal to the sum of the negative eigenvalues
of $\rho^{T_A}$. The realignment, or cross-norm \cite{Rudolph:02,
Chen/Wu} can detect entanglement for which the negativity fails.
Similar to the negativity, it is defined in terms of a relabeling
\begin{equation}
{{{\mathcal{R}}}}(\rho)_{ij,kl}=\rho _{ik,jl},
\end{equation}
which results in the following measure
\begin{equation}
{{{\mathcal{R}}}}=\max {\lbrace {\frac{\Vert {{{\mathcal{R}}}}(\rho)
\Vert -1}{2}, 0} \rbrace }.
\end{equation}
For simplicity, we refer to the quantity ${{{\mathcal{R}}}}$ as the
realignment. We take the maximum in the above equation since the
first term in the bracket can be negative for some states. Note that
${{{\mathcal{N}}}}= 0$ is necessary for separability of any
bipartite states and is also sufficient for $2\otimes 2$ and $2
\otimes 3$ states.  But the realignment ${{{\mathcal{R}}}}= 0$ is a
necessary condition of separability for all bipartite states.

By using the monogamy of concurrence (\ref{a}), the monogamy of
negativity and realignment has been proven
\cite{Ou/Fan:07,Fan/etal:07}. Thus, for any pure three-qubit states
$\rho_{ABC}$, we have
\begin{equation}  \label{g1}
{{{\mathcal{N}}}}^2_{AB}+{{{\mathcal{N}}}}_{AC}^2\leq
{{{\mathcal{N}}}}^2_{A:BC},
\end{equation}
and
\begin{equation}  \label{g2}
{{{\mathcal{R}}}}^2_{AB}+{{{\mathcal{R}}}}_{AC}^2\leq
{{{\mathcal{R}}}}^2_{A:BC}.
\end{equation}

The objective of this paper is to provide a set of general and
computable constraints on entanglement sharing for multipartite
quantum states. To do this, we first note that the general monogamy
inequalities for pure multi-qubit states $\rho_{A_1A_2...A_n}$ in
terms of concurrence, negativity, and realignment. The concurrence
satisfies
\begin{equation}\label{eq:ineq1}
{{{\mathcal{C}}}}^2_{A_1A_2}+{{{\mathcal{C}}}}^2_{A_1A_3}+\cdot\cdot\cdot+{{{\mathcal{C}}}}^2_{A_1A_n}
\leq {{{\mathcal{C}}}}^2_{A_1:A_2A_3...A_n},
\end{equation}
the negativity satisfies
\begin{equation}\label{eq:ineq2}
{{{\mathcal{N}}}}^2_{A_1A_2}+{{{\mathcal{N}}}}^2_{A_1A_3}+\cdot\cdot\cdot+{{{\mathcal{N}}}}^2_{A_1A_n}
\leq {{{\mathcal{N}}}}^2_{A_1:A_2A_3...A_n},
\end{equation}
and the realignment also satisfies
\begin{equation}\label{eq:ineq3}
{{{\mathcal{R}}}}^2_{A_1A_2}+{{{\mathcal{R}}}}^2_{A_1A_3}+\cdot\cdot\cdot
               +{{{\mathcal{R}}}}^2_{A_1A_n}
      \leq {{{\mathcal{R}}}}^2_{A_1:A_2A_3...A_n},
\end{equation}
where ${{{\mathcal{C}}}}_{A_1A_i}$, ${{{\mathcal{N}}}}_{A_1A_i}$,
and ${{{\mathcal{R}}}}_{A_1A_i}$ denote two-qubit quantities and
${{{\mathcal{C}}}}^2_{A_1:A_2A_3...A_n}$,
${{{\mathcal{N}}}}^2_{A_1:A_2A_3...A_n}$, and
${{{\mathcal{R}}}}^2_{A_1:A_2A_3...A_n}$ are the bipartite
entanglement measured by the respective quantities across the
(bi)partition $A_1:A_2A_3...A_n$. The detailed proof of the
inequalities (\ref{eq:ineq1}) and (\ref{eq:ineq2}) can be found in
\cite{Coffman/Kundu/Wootters,Osborne/Verstraete:96, Ou/Fan:07}. The
verification of the inequality (\ref{eq:ineq3}) follows immediately
from the work of \cite{Ou/Fan:07, Fan/etal:07}, so is omitted here.

\section{Computable monogamy inequality in terms of negativity and
  realignment}
\noindent It should be emphasized that the inequalities
(\ref{eq:ineq1}-\ref{eq:ineq3}) are limited to bipartite subsystems,
i.e., $A_i$ is a single qubit, and do not provide the extent to
which the general monogamy inequality applies. This is important
because it has been shown that the monogamy of concurrence does not
generally hold for higher-dimensional quantum states \cite{Ou:07}.
In view of the difficulty of analytically calculating entanglement
of bipartite states with dimension greater than 2, for pure
multi-qubit states, we propose a more general monogamy constraint on
the corresponding distribution of entanglement. The most important
thing is that it is computable, as shown in the following theorem,
one of two main results of this paper.

{\it Theorem 1}: For pure multi-qubit states $\rho_{ABC...DE}$ where
$A, B, C, ..., D$ stand for single qubits, respectively, while  $E$
a collection of any number of qubits, i.e., $E=(E_1E_2...E_l)$ where
$E_1, E_2, ..., E_l$ are qubits, the negativity and realignment
satisfy the following inequalities
\begin{equation}\label{p4}
{{\mathcal{N}}}^2_{AB}+{{\mathcal{N}}}^2_{AC} +\cdot\cdot\cdot+
  {{\mathcal{N}}}^2_{AD} + {{\mathcal{N}}}^2_{AE}\leq {{\mathcal{N}}}^2_{A:BC...DE},
\end{equation}
and
\begin{equation}\label{p5}
{{\mathcal{R}}}^2_{AB}+{{\mathcal{R}}}^2_{AC} +\cdot\cdot\cdot+
  {{\mathcal{R}}}^2_{AD} + {{\mathcal{R}}}^2_{AE}\leq
  {{\mathcal{R}}}^2_{A:BC...DE}.
\end{equation}

{\it Proof}: The proof is straightforward and stems from the fact
that both the negativity and the realignment are lower bounds of the
concurrence.  For a pure bipartite state $\rho_{F_1F_2} =
|\Psi_{F_1F_2}\rangle \langle\Psi_{F_1F_2}| $ in $2 \otimes d $
dimensions, the system has the Schmidt form
\begin{equation}\label{p9}
{|\Psi_{F_1F_2}\rangle}=\sqrt{\lambda_1}|e_1
f_1\rangle+\sqrt{\lambda_2}|e_2 f_2\rangle,
\end{equation}
where $\sqrt{\lambda_i}$ are Schmidt coefficients and $|e_i\rangle$
and $|f_i\rangle$ are orthonormal bases for ${\mathcal{H}}_{F_1}$
and ${\mathcal{H}}_{F_2}$, respectively. It is shown that
\cite{Chen/etal:05,Ou/Fan:07}
\begin{equation}
{{\mathcal{C}}}_{F_1F_2}=2{{\mathcal{N}}}_{F_1F_2}=2{{\mathcal{R}}}_{F_1F_2}=2\sqrt{\lambda_1\lambda_2}.
\end{equation}
In a similar way, we have
\begin{equation}\label{y1}
{{\mathcal{C}}}_{A:BC...DE}=2{{\mathcal{N}}}_{A:BC...DE}=2{{\mathcal{R}}}_{A:BC...DE}.
\end{equation}
Now if $\rho_{F_1F_2}$ is a mixed state, then \cite{Chen/etal:05}
\begin{equation}\label{y2}
2{{\mathcal{N}}}_{F_1F_2}\leq {{\mathcal{C}}}_{F_1F_2},
\end{equation}
and
\begin{equation}\label{y3}
2{{\mathcal{R}}}_{F_1F_2}\leq {{\mathcal{C}}}_{F_1F_2}.
\end{equation}
Note that if $F_1$ is not two-dimensional, some extra constant
factors will appear in the expressions on the left-hand sides of
Eqs.~(\ref{y2}) and (\ref{y3}).

For either pure or mixed states $\rho_{GHK}$ in $2\otimes 2\otimes
2^{n-2}$ dimensions we have \cite{Osborne/Verstraete:96}
\begin{equation}  \label{y4}
{{\mathcal{C}}}^2_{GH}+{{\mathcal{C}}}_{GK}^2\leq
{{\mathcal{C}}}^2_{G:HK}.
\end{equation}
In order to obtain our desired result, we partition $K$ into two
parts, i.e., $K_1$ and $K_2$. $K_1$ is a single qubit while $K_2$ is
a collection of $n-3$ qubits. Therefore, according to
Eq.~(\ref{y4}), the following inequality holds
\begin{equation}  \label{y55}
{{\mathcal{C}}}^2_{GH}+{{\mathcal{C}}}_{GK_1}^2+{{\mathcal{C}}}_{GK_2}^2\leq
{{\mathcal{C}}}^2_{G:HK}.
\end{equation}
Analogously, successively applying Eqs.~(\ref{y4}) and (\ref{y55})
to the pure state $\rho_{ABC...DE}$ gives
\begin{equation}\label{y5}
{{\mathcal{C}}}^2_{AB}+{{\mathcal{C}}}^2_{AC} +\cdot\cdot\cdot+
  {{\mathcal{C}}}^2_{AD} + {{\mathcal{C}}}^2_{AE}\leq {{\mathcal{C}}}^2_{A:BC...DE},
\end{equation}
such that the proofs of (\ref{p4}) and (\ref{p5}) can be finished by
considering (\ref{y1}-\ref{y3}) and (\ref{y5}).

We remark that {\it Theorem 1} would be much more general if $A, B,
C,...D$ contained an arbitrary number of qubits, respectively.
However, since it is much easier to calculate ${{\mathcal{N}}}_{AE}$
and ${{\mathcal{R}}}_{AE}$ than ${{\mathcal{C}}}_{AE}$, we believe
these may provide a certain quantification of the correlations,
which could be valuable in quantum cryptographic protocols and among
other uses.
\vspace*{4pt}   
\begin{table}[tbp]
\tcaption{Values of the residual entanglement defined by the
$\tau_{ABC}$ and $\pi_{ABC}$ different classes.} \label{table}
\centerline{\footnotesize\smalllineskip
\begin{tabular}{cccccccccccccccccccccccccccccc}
\hline\hline Class && && & & &&
 $\tau_{ABC}$
 & & &
&& & &&
 $\pi_{ABC}$  \\ \hline A-B-C
 && & & &&  & & 0
& && & & && & 0
\\ \hline A-BC  && && & & & &
 0 & && && && &0     \\ \hline B-AC  && && & & & & 0
 & && & & && &0     \\ \hline C-AB && && & & & & 0          & && && & & &  0     \\
\hline W     & & && & & && 0 & && && & & & $>0$
\\ \hline GHZ   & & && & & && $>0$   & && && && &$>0$
\\\hline\hline
\end{tabular}}
\end{table}

Note that our result is not limited to pure multi-qubit states. For
mixed states these monogamy inequalities also hold when the
right-hand side is minimized over all pure-state decompositions as
in (\ref{eq:minconc}).  This can be done using the convexity of
negativity and realignment.  However, the quantitative evaluation
then becomes intractable for large systems. By decomposing a
higher-dimensional Hilbert space into lower-dimensional spaces,
useful bounds on entanglement between different parties can be
obtained.

For a tripartite state, the smallest Hilbert space is $2 \otimes 2
\otimes 2$ dimensional.  The bound for the entanglement gives some
indications of the distribution of entanglement of the whole state.
For example, for either pure or mixed three-qubit states
$\rho_{ABC}$, the concurrence between $A$ and $B$ is limited by the
one between $C$ and $AB$ \cite{Huang/Zhu:08}, i.e.,
\begin{equation}\label{p6}
{{\mathcal{C}}}_{AB} \leq
\frac{1}{2}\left(1+\sqrt{1-{{{\mathcal{C}}}}^2_{C:AB}}\right).
\end{equation}

Given Eqs.~(\ref{y2}), (\ref{y3}), and (\ref{p6}) we can also state
the following theorem concerning the monogamy of entanglement as
measured by the negativity and realignment.

{\it Theorem 2}:  For any three-qubit state $\rho_{ABC}$, the
entanglement between $A$ and $B$, and the one between $C$ and $AB$
satisfy
\begin{equation}\label{p8}
{{\mathcal{N}}}_{AB} \leq
\frac{1}{4}\left(1+\sqrt{1-4{{\mathcal{N}}}^2_{C:AB}}\right),
\end{equation}
and
\begin{equation}\label{p82}
{{\mathcal{R}}}_{AB} \leq
\frac{1}{4}\left(1+\sqrt{1-4{{\mathcal{R}}}^2_{C:AB}}\right).
\end{equation}

Again, we emphasize the computability of these results.


\section{Application and discussion}

\noindent In this section, several examples are discussed which show
the utility of the monogamy inequalities established above.

\subsection{Examples}
\noindent To compare various classes of entangled states, we recall
the definition of the residual entanglement
\begin{equation}
\label{pp}
\pi_{ABC}={\mathcal{N}}^2_{A:BC}-{\mathcal{N}}^2_{AB}-{\mathcal{N}}^2_{AC}.
\end{equation}
This quantity is greater than zero for pure three-qubit states
$\rho_{ABC}$, for the $W$-class and $GHZ$-class \cite{Ou/Fan:07},
while the quantification of entanglement given by the three-tangle
in Eq.~(\ref{22}) is greater than zero only for the $GHZ$-class
\cite{Dur/etal:00}.  Although it has been demonstrated in
Ref.~\cite{Ou/Fan:07}, we provide a comparison of the inequalities
in Table~\ref{table}. Note that all pure three-qubit states can be
sorted into six classes through stochastic local operation and
classical communication (SLOCC) \cite{Dur/etal:00}. (1)
$A$--$B$--$C$ class including product states; (2) $A$--$BC$, (3)
$B$--$AC$, and (4) $C$--$AB$ classes including only bipartite
entangled states; (5) $W$ and  (6) GHZ classes including so-called
genuine tripartite entangled states. It is evident that one can
distinguish the two classes of fully entangled states with the help
of $\tau_{ABC}$ and $\pi_{ABC}$. Furthermore, we have numerically
verified that the monogamy inequality of realignment,
Eq.~(\ref{g2}), is also strict for pure three-qubit states.  Given
this, we have the following conjecture.


\subsection{A Conjecture}
\noindent {\it Conjecture}: For a pure multipartite state $\rho_{ABCD...}$, if
the particle $A$ is entangled with at least two other particles, the
monogamy inequalities (\ref{p4}-\ref{p5}) are strict.

\begin{figure}[tbp]
\begin{center}
\includegraphics[scale=0.75,angle=0]{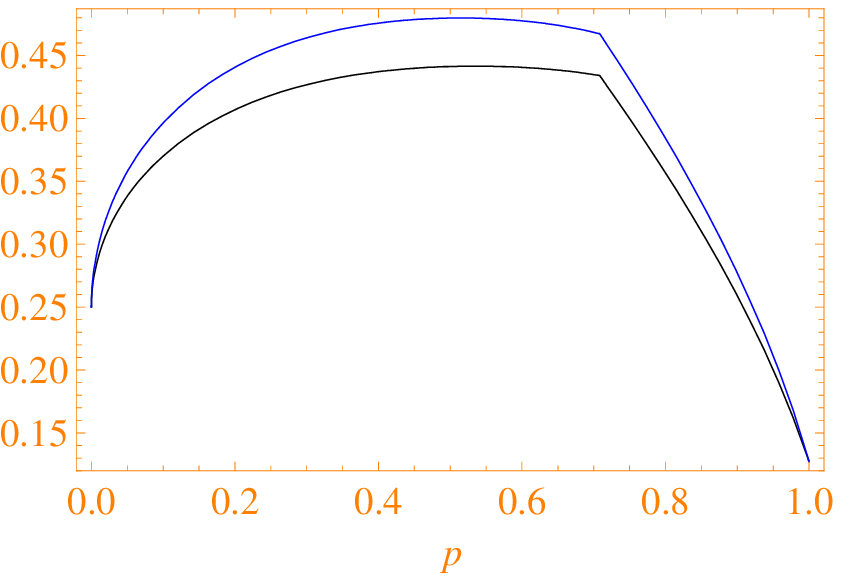} %
\fcaption{ Plot of $\tau_{{\mathcal{N}}}$ (lower curve) and
$\tau_{{\mathcal{R}}} $(upper curve) for the states (\ref{po}) as a
function of $p$.} \label{fig:NRcomparison}
\end{center}
\end{figure}

On the other hand, the inequalities Eq.~(\ref{p8}) and
Eq.~(\ref{p82}) have similar properties for pure states. For a pure
state $\rho_{ABC}$ in the $GHZ$-class, it is obvious that the two
inequalities are strict since
${\mathcal{N}}_{AB}={\mathcal{R}}_{AB}=0$ for this class while the
right-hand sides of Eq.~(\ref{p8}) and Eq.~(\ref{p82}) are not less
than 0.25.  Some $W$-class states saturate the inequality
Eq.~(\ref{p6}), and since it has been shown that
$2{\mathcal{N}}_{AB} < {\mathcal{C}}_{AB}$ \cite{Ou/Fan:07}, the
inequality Eq.~(\ref{p8}) is also strict for this class. Likewise,
the inequality Eq.~(\ref{p82}) is strict for the $W$-class.  As for
the other four classes, only for the following entangled states
\begin{equation}
|\phi\rangle_{AB}\otimes|\phi\rangle_{C}= U_A\otimes U_{B}\otimes I
\frac{1}{\sqrt{2}}(|00\rangle+|11\rangle)_{AB}\otimes|\phi\rangle_{C},
\end{equation}
where $U_A, U_B \in U(2)$, do the inequalities Eq.~(\ref{p8}) and
Eq.~(\ref{p82}) become equalities.

Now let us investigate the rank-2 mixed three-qubit states
\begin{equation}\label{po}
\rho{(p)}=p|W\rangle\langle W|+(1-p)|GHZ\rangle\langle GHZ|,
\end{equation}
with $p \in [0,1]$.  We define $\tau_{{\mathcal{N}}}$ and
$\tau_{\mathcal{R}}$ to be the difference between the left-hand side
and right-hand side of Eq.~(\ref{p8}) and Eq.~(\ref{p82})
respectively, i.e.,
\begin{equation}\label{p81}
\tau_{\mathcal{N}}=
\frac{1}{4}\left(1+\sqrt{1-4{\mathcal{N}}^2_{C:AB}}\right)-{\mathcal{N}}_{AB},
\end{equation}
and
\begin{equation}\label{p91}
\tau_{\mathcal{R}}=\frac{1}{4}\left(1+\sqrt{1-4{\mathcal{R}}^2_{C:AB}}\right)-{\mathcal{R}}_{AB}.
\end{equation}
We plot these in Fig.~\ref{fig:NRcomparison} for the states given in
Eq.~(\ref{po}). From Fig.~\ref{fig:NRcomparison}, we can see that
the monogamy of negativity Eq.~(\ref{p81}) and realignment
Eq.~(\ref{p91}) provide different constraints for these states, in
this sense, generally they are inequivalent.  For some states
however, they are complementary. In contrast to Eq.~(\ref{p81}) and
Eq.~(\ref{p91}), for mixed states the monogamy constraints for
distributed entanglement Eq.~(\ref{ppp}) and Eq.~(\ref{p6}) are very
difficult to calculate.

\subsection{Experimental Determination}

\noindent For applications of this work, we provide a direct route
to experimental confirmation of these inequalities for sets of
qubits. We do this by noting that the components of the polarization
vector (a.k.a the generalized coherence vector, or Bloch vector
representation \cite{Mahler:book,Jakob:01,Byrd/Khaneja:03,Kimura})
are experimentally measurable quantities.  We note that several
important examples of the concurrence \cite{Byrd/Khaneja:03}, the
realignment \cite{Rudolph:02}, and the negativity
\cite{Byrd/Brennen:08}, have already been expressed in terms of the
polarization vector.

To be specific, the square of the concurrence for any two two-state
subsystems, $A$, $B$ can be expressed in terms of the square roots
of the eigenvalues of the matrix
\begin {equation} M =
\rho_{AB}\tilde{\rho}_{AB},
\end{equation}
 which we will let be
$\lambda_1,\lambda_2,\lambda_3,\lambda_4$ with $\lambda_i\geq
\lambda_{i+1}$.  The square of the concurrence for two subsystems
$A$ and $B$ is
\begin{eqnarray}
{\cal C}_{AB}^2 &=& (\lambda_1-\lambda_2-\lambda_3-\lambda_4)^2 \nonumber \\
                &=& \lambda_1^2 + \lambda_2^2 + \lambda_3^2
                +\lambda^2_4 \nonumber \\
                 &&  -2\lambda_1\lambda_2-2\lambda_1\lambda_3-2\lambda_1\lambda_4 \nonumber \\
                 &&  + 2\lambda_2\lambda_3 +2\lambda_2\lambda_4+2\lambda_3\lambda_4.
\end{eqnarray}
Note that the following inequality holds which is useful for the
inequalities provided later: \begin{equation} \lambda_1\lambda_2+
\lambda_1\lambda_3+\lambda_1\lambda_4 \geq \lambda_2\lambda_3+
\lambda_2\lambda_4+\lambda_3\lambda_4.
\end {equation} In fact each term on the left is, taking the terms
in order, greater than those on the right.

Now, dropping the negative terms, and adding a few \begin{equation}
C_{AB} \leq \texttt{Tr} {(M)} +2[S_2(M)]^{1/2}, \end {equation}
where $S_2(M) =(1/2)[(\texttt{Tr}{(M)})^2 -\texttt{Tr}(M^2)]$
\cite{Byrd/Khaneja:03}. Dropping the positive terms, and subtracting
a few, we obtain
\begin{equation} C_{AB} \geq \texttt{Tr}(M) - 2[S_2(M)]^{1/2}. \end
{equation} Clearly this latter inequality is only meaningful when
both the concurrence and the right-hand side are both greater than
zero.

Now $\rho_{AB}$ and $\tilde{\rho}_{AB}$, for qubits, can be
expressed in terms of the polarization vector as \begin {equation}
\rho_{AB} = \frac{1}{4}\left(I +\vec{n}_A\cdot\vec{\sigma}\otimes I
+ I \otimes \vec{n}_B\cdot\vec{\sigma} +
\sum_{\alpha,\beta}c_{\alpha\beta}\sigma_\alpha\otimes\sigma_\beta\right),
\end {equation} and \begin {equation} \tilde{\rho}_{AB} =
\frac{1}{4}\left(I -\vec{n}_A\cdot \vec{\sigma}\otimes I - I\otimes
\vec{n}_B\cdot\vec{\sigma} +
\sum_{\alpha,\beta}c_{\alpha\beta}\sigma_\alpha\otimes\sigma_\beta\right).
\end {equation}

One can multiply these two and to obtain $M$ and thus all bounds on
the concurrence given in this article in terms of experimentally
available quantities.  The required measurements only require, at
most, sets of two-body interactions.

In the case that the density operator under consideration has
single-particle density operators which are completely mixed, the
bounds simplify dramatically.  The matrix $\tilde{\rho}$ becomes
equal to $\rho$, \begin {equation} \tilde{\rho}_{AB} = \rho_{AB} =
\frac{1}{4}\left(I +
\sum_{\alpha,\beta}c_{\alpha\beta}\sigma_\alpha\otimes\sigma_\beta\right).
\end {equation} The bounds above can be computed and measured by well-known
methods \cite{Brun:04}.

Furthermore, for the two-qubit case that the local density operators
are completely mixed, the realignment, or cross-norm, can also be
readily computed in terms of measurable quantities without relying
on bounds. Since the entanglement measures are invariant under local
unitary transformations, local unitary transformations can be used
to diagonalize the correlation matrix $c_{\alpha\beta}$, to get
$c_{\alpha\alpha}$ and the realignment gives \begin {equation} {\cal
R}(\rho) =\frac{1}{2}\left(1+\sum_\alpha
\left|c_{\alpha\alpha}\right|\right). \end {equation}

These are themselves measurable quantities, as are the
$c_{\alpha\beta}$ before diagonalization. Similarly the negativity
for the two-qubit state can be found in terms of the correlation
matrix components by direct computation, \begin {eqnarray} {\cal
N}(\rho_{AB}) &=& \frac{1}{2}\left(
  \left|\frac{1}{4}+c_{33}+c_{11}-c_{22}\right|\right. \nonumber \\
                 && + \left|\frac{1}{4}+c_{33}-c_{11}+c_{22}\right|
                   + \left|\frac{1}{4}-c_{33}+c_{11}+c_{22}\right|
                   \nonumber \\
                 &&\left. +\left|\frac{1}{4}-c_{33}-c_{11}-c_{22}\right| \right).
\end {eqnarray}

These expressions provide a direct route to experimental
determination of the bounds provided in this article using the
two-particle correlations.


\section{Conclusions}

\noindent

Here we have derived monogamy inequalities in terms of negativity
and realignment. Since these two measures of entanglement are
computable, these inequalities provide a quantitative evaluation of
constraints on entanglement-sharing for multipartite states. Most
importantly, Eq.(\ref{p4}) and Eq.(\ref{p5}) provide inequalities
which are not necessarily limited to bipartite systems and are able
to give more information about the possibilities for distributing
entanglement. The importance of these also lie in the fact they
allow us to investigate the distribution of entanglement  based on
different partitions of many-qubit systems in condensed matter
physics. We also derive computable monogamy inequalities for any
three-qubit states. Furthermore, experimental determination of these
bounds may be accomplished given the expressions here. In the near
future, studying the entanglement in three-qubit states as done in
\cite{u11} through the results in this paper is worthwhile, since
different monogamy inequalities can lead to different residual
entanglement. We believe our work will greatly aid in the
understanding of the physical implications of these inequalities.

The most appealing feature of Eq.~(\ref{p8}) and Eq.~(\ref{p82}) is
that these quantities are {\it computable} entanglement-sharing
constraints for mixed three-qubit states.


\nonumsection{Acknowledgements} \noindent
This material is based upon
work supported by the National Science Foundation under Grant No.
0545798. We thank Shao-Ming Fei and Heng Fan for valuable
discussions.


\nonumsection{References}

\end{document}